\documentclass[sigconf, nonacm]{acmart}
\AtBeginDocument{%
  }

\renewcommand\footnotetextcopyrightpermission[1]{} 
\usepackage{soul}
\usepackage{algorithm} 
\usepackage{algpseudocode} 
\usepackage{tabularx}
\usepackage{pifont}
\usepackage{multirow}
\usepackage{colortbl} 
\usepackage{caption}
\usepackage{float}
\usepackage[utf8]{inputenc}
\usepackage{todonotes}

\begin{document}
\title{Light-Bound Transformers: Hardware-Anchored Robustness for Silicon-Photonic Computer Vision Systems\vspace{-2mm}}

\author{Xuming Chen$^{*,1}$, Deniz Najafi$^{*,2}$, Chengwei Zhou$^{1}$, Pietro Mercati$^3$, Arman Roohi$^4$\\ Mohsen Imani$^{5}$, Mahdi Nikdast$^{6}$, Shaahin Angizi$^2$, and Gourav Datta$^{1}$  \vspace{0.3em}\\ \small $^1$Case Western Reserve University, USA $^2$New Jersey Institute of Technology, USA 
$^3$Intel Corporation, USA  \\
$^4$University of Illinois Chicago, USA
$^5$University of California, Irvine, USA
$^6$Colorado State University, USA \\
$^{*}$ Equal Contributions
 \vspace{-0.3em}
\\}


\begin{abstract}
Deploying Vision Transformers (ViTs) on near-sensor analog accelerators demands training pipelines that are explicitly aligned with device-level noise and energy constraints. We introduce a compact framework for silicon-photonic execution of ViTs that integrates {measured hardware noise}, {robust attention training}, and an {energy-aware processing flow}. We first characterize bank-level noise in microring-resonator (MR) arrays, including fabrication variation, thermal drift, and amplitude noise, and convert these measurements into closed-form, activation-dependent variance proxies for attention logits and feed-forward activations. Using these proxies, we develop \emph{Chance-Constrained Training} (CCT), which enforces variance-normalized logit margins to bound attention rank flips, and a \emph{noise-aware LayerNorm} that stabilizes feature statistics without changing the optical schedule.  These components yield a practical ``measure $\rightarrow$ model $\rightarrow$ train $\rightarrow$ run'' pipeline that optimizes accuracy under noise while respecting system energy limits. Hardware-in-the-loop experiments with MR photonic banks show that our approach restores near-clean accuracy under realistic noise budgets, with no in-situ learning or additional optical MACs.
\vspace{-4mm}

\end{abstract}



\maketitle
\pagestyle{plain}
 \vspace{-2.4em} 

\section{Introduction}

Transformer architectures have become the default backbone for modern vision tasks, from image classification to dense prediction, owing to their scalable receptive fields and data-driven inductive biases \cite{dosovitskiy2021an}. Vision Transformers (ViTs) replace convolutional correlation with learned self-attention, repeatedly forming content-dependent matrix products such as $QK^\top$ and $AV$ across layers. While this structure excels on server-grade processors, its repeated matrix–vector operations and quadratic token interactions stress the energy and bandwidth budgets of edge systems~\cite{liu2025lawcat}. This tension has motivated a surge of interest in both digital~\cite{vita} and analog accelerators~\cite{ambrogio2018equivalent,rasch2023hardware,dong2026inmemoryadcbasednonlinearactivation}, including electronic in-memory computing (IMC) and silicon–photonic multiply–accumulate fabrics, that promise orders-of-magnitude improvements in bandwidth density and energy per MAC by collocating weights and computation. In photonics in particular, integrated interferometer meshes and microring-resonator (MR) banks can realize large linear transforms with low latency and high throughput \cite{shen2017deep} \cite{microring} \cite{microring2},  provided that non-idealities such as fabrication-induced detuning, thermal drift, and source amplitude noise are managed \cite{bogaerts2012silicon,padmaraju2014resolving}.

\begin{figure}[t] \vspace{-.4em}
\centering
\includegraphics [width=0.99\linewidth,]{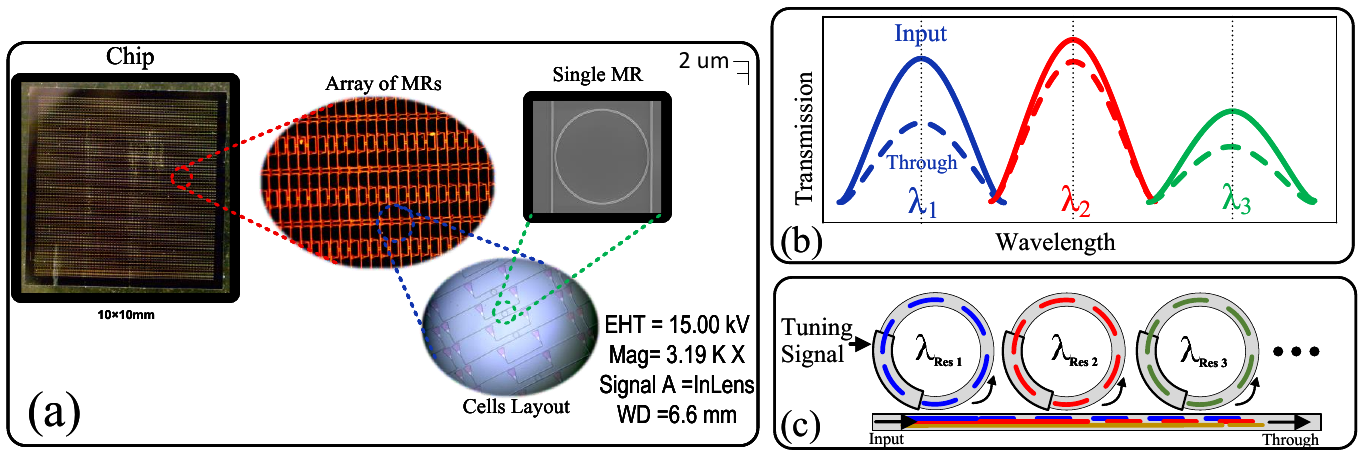}
\vspace{-1em}
\caption{\small (a) Fabricated SiPh MR array with >200 identical MR cells (SEM shown).
(b) Input and through-port spectra after parameter imprinting via tuning the MR resonance.
(c) Multiple MRs in one arm imprint weight values onto the input signal at different wavelengths.} 
\vspace{-2em}
\label{mr}
\end{figure}

A central obstacle to deploying ViTs on such analog substrates is \emph{noise-aware learning at the right locus}. Conventional fine-tuning with i.i.d.\ Gaussian perturbations improves average-case resilience but does not target the pairwise logit orderings that govern attention routing, nor does it leverage the structure of device noise observed on real hardware (e.g., per-bank variance/covariance on MR arrays shown in Fig. \ref{mr} or read/program variance in IMC crossbars). Moreover, standard normalization (e.g., LayerNorm \cite{ba2016layer}) stabilizes hidden states but is agnostic to the measured, bank-level statistics that determine how activation energy couples into logit variance. This creates a gap between device-level characterization and algorithm-level robustness, leading to either over-approximation or under-modeling of noise and brittle deployment behavior.

We propose a measurement-driven methodology for robust ViT deployment on analog accelerators, demonstrated with silicon-photonic MR banks, that (i) maps device noise sources to per-logit variance proxies, (ii) introduces Chance-Constrained Training (CCT) to directly bound the probability of attention flips via variance-normalized logit gaps, and (iii) adds a noise-aware LayerNorm to stabilize activations under hardware noise. All variance proxies are computed analytically per forward pass, making the approach practical for large ViTs, and together these tools translate device statistics into differentiable, energy-efficient training and inference objectives.
We validate the approach using benchtop measurements and hardware-in-the-loop emulation of MR banks. Across ImageNet-scale ViT models, CCT and noise-aware normalization reliably recover clean accuracy under realistic noise, while the inference flow exposes accuracy–energy tradeoffs unique to analog execution. Unlike generic randomized smoothing \cite{cohen2019rs}, our guarantees are \emph{hardware-relevant}, directly bounding attention flip probabilities under measured bank-level noise. Our method is complementary to advances in analog IMC robustness \cite{rasch2023hardware,ambrogio2018equivalent}, offering a principled path toward robust, energy-efficient ViTs on photonic and IMC substrates by aligning learning objectives with device statistics and integrating noise into the inference pipeline.

\vspace{-2mm}
\section{Background}\vspace{-0.2em}
\noindent\textbf{Vision Transformers.}
The Vision Transformer (ViT)~\cite{dosovitskiy2021an} adapts transformer encoder architectures from BERT~\cite{devlin2019bertpretrainingdeepbidirectional} for vision tasks. The input is split into $n$ patches and embedded into an $n \times d_m$ matrix. Each of the $L$ encoder blocks consists of Multi-Head Self-Attention (MHSA) and Feed-Forward Network (FFN) modules with layer normalization and residual connections. MHSA uses $h$ heads with query, key, and value projections ($W_Q, W_K, W_V$), computing attention as $\texttt{softmax}\left(\frac{QK^T}{\sqrt{d_k}}\right)V$. The outputs are concatenated and passed through the FFN to model complex relationships in the input.




\noindent\textbf{MicroRing Resonators and SiPh Acceleration.} SiPh-based accelerators offer high bandwidth and address fan-in/fan-out challenges for DNN and vision tasks~\cite{sunny2021crosslight,liu2019holylight,zokaee2020lightbulb,xu202111,shiflett2021albireo}. They can be broadly categorized into coherent designs using a single wavelength~\cite{zhao2019hardware} and non-coherent designs leveraging multiple wavelengths for parallelism~\cite{sunny2021crosslight,sunny2021robin}. In non-coherent systems, microring resonators (MRs) dynamically modulate light intensity to encode inputs and/or weights~\cite{sunny2021crosslight,sunny2021robin,morsali2024lightator}. MRs enable efficient MAC operations by tuning resonant wavelengths, given by $\lambda_{res}=\frac{n_{eff}L}{m}$~\cite{bogaerts2012silicon}. Prior MR-based accelerators include LightBulb~\cite{zokaee2020lightbulb}, which accelerates binarized CNNs but incurs high ADC overhead; ROBIN and CrossLight~\cite{sunny2021robin,sunny2021crosslight}, which improve efficiency with low-bit weights but still rely on costly data converters; and Lightator~\cite{morsali2024lightator}, which targets near-sensor DNN acceleration with compressive sensing. More recently, Opto-ViT~\cite{optovit} introduces a hybrid electronic-photonic ViT accelerator that leverages WDM-enabled MR cores for matrix multiplications and employs region-of-interest masking to reduce redundant computation, achieving high energy efficiency.

\noindent\textbf{Noise-aware Training.}
Analog photonic neural systems are vulnerable to high error rates and computational inaccuracy due to analog distortions and pervasive optoelectronic noise~\cite{ohno2022si,moon2019enhancing,joshi2020accurate,hu2016dot,photonicnoise}. While noise modeling for SiPh designs is still limited, electronic IMC resistive crossbars have been more extensively studied and mitigated using offline noise-aware training methods \cite{victor2025memory,mao2022experimentally,yang2021multi,noiseamin,mirzafab}. These include injecting stochastic perturbations into inputs~\cite{bishop1995training}, weights~\cite{blundell2015weight}, and activations~\cite{rekhi2019analog}. Such strategies have improved inference robustness in resistive crossbar architectures, but mainly for discriminative models, and results for SiPh-based systems remain limited. Diffractive optical neural networks have used parametric randomness to increase tolerance to optical imperfections~\cite{mengu2020misalignment}. Notably, some approaches leverage photonic hardware's inherent analog noise as a resource in machine learning algorithms~\cite{wu2022harnessing}.


\vspace{-0.5em} 

\begin{figure}[t] \vspace{-2em}
\centering
\includegraphics [width=0.99 \linewidth]{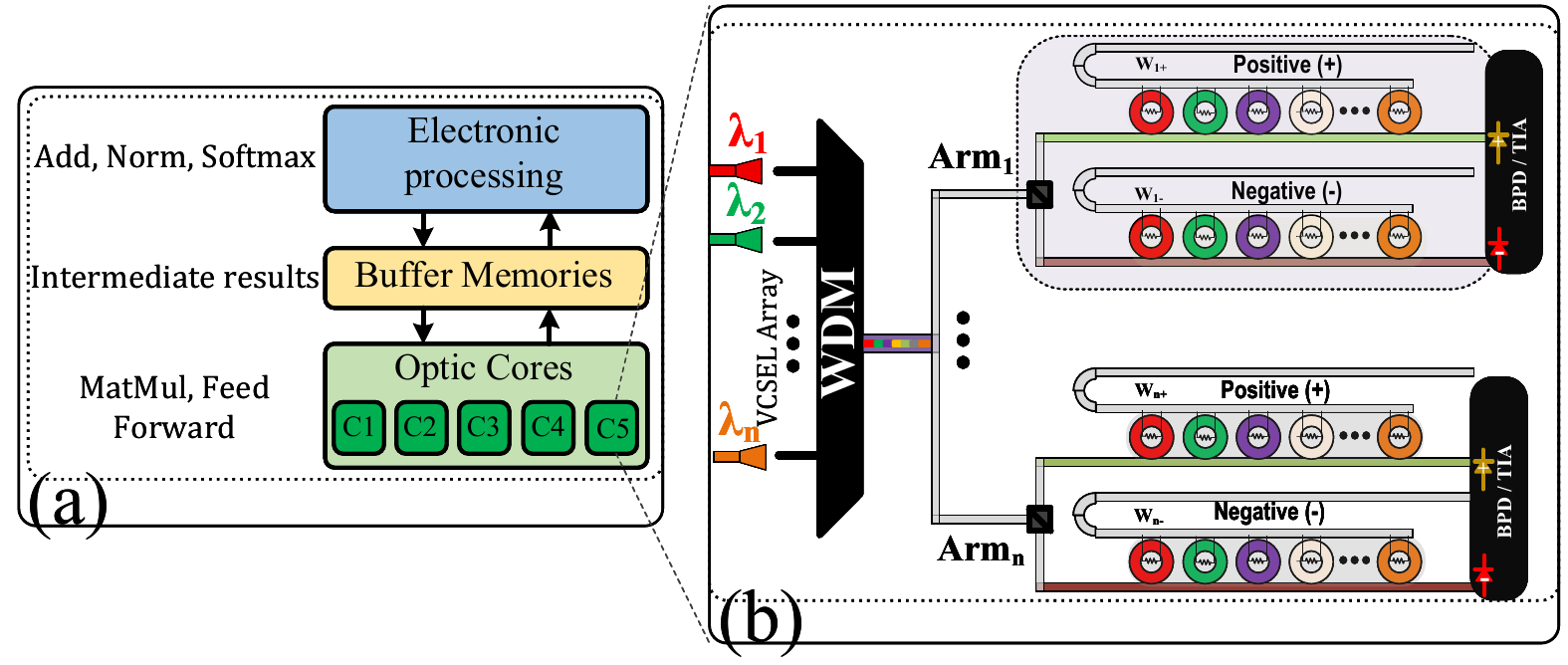}
\vspace{-1.8em}
\caption{\small Overview of the proposed noisy under-test architecture.} 
\vspace{-2.6em}
\label{arch}
\end{figure}

\section{Architecture and Noise Characterization}
\vspace{-1mm}
\subsection{Proposed Under-test Architecture}
 The proposed under-test architecture (Fig. \ref{arch} (a))  employs a hybrid electronic–photonic architecture comprising optical and electronic processing blocks with a shared buffer memory. The optical block, comprising five optical cores, implements the most computation-intensive transformer primitives—including the matrix-matrix multiplication (MatMuls) in MHSA, FFN, and embedding layers—while leveraging wavelength-division multiplexing (WDM) to orchestrate highly parallel vector–vector and matrix–matrix multiplications across multiple wavelengths and cycles. 
 As shown in Fig. \ref{arch}(b), each core integrates 64 waveguide arms, each hosting 32 MRs allocated to be tuned by positive and negative weights across 32 wavelength channels, along with multiplexers, and driver/modulator circuits and Vertical-Cavity Surface-Emitting Laser (VCSELs) arrays which directly modulate input data into light intensity for optical MAC. 
Meanwhile, the electronic block performs non-linear functions—Softmax, GELU, normalization, and additions—enabling efficient integration of optical computation with precise digital control. Details of MatMul mapping and implementation are discussed in the next subsection.

 \begin{figure}[t] \vspace{-2em}
\centering
\includegraphics [width=1.01\linewidth]{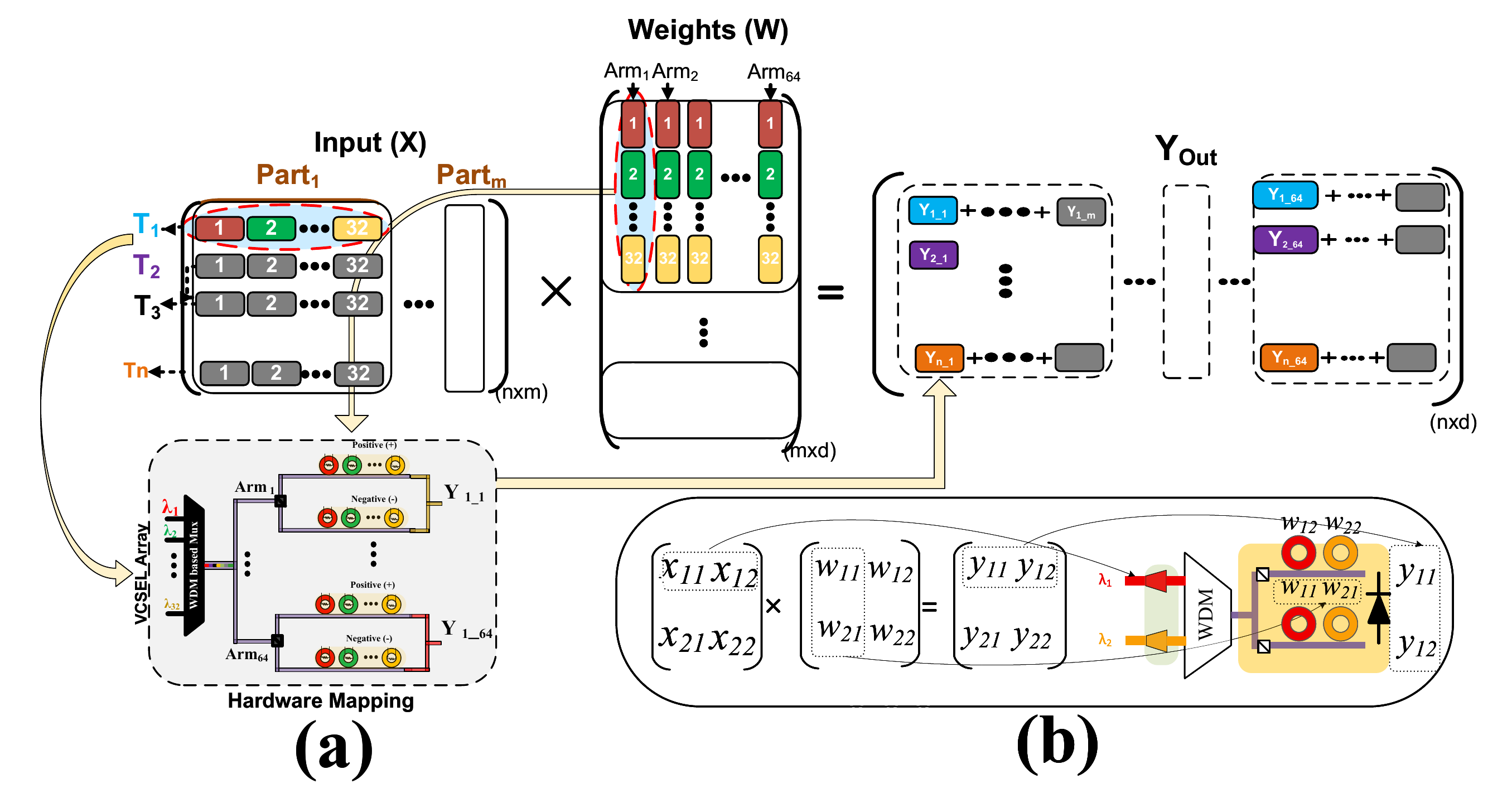}
\vspace{-2.5em}
\caption{\small (a) Matrix splitting and mapping methodology, (b) Optical matrix-matrix multiplication.} 
\vspace{-2.em}
\label{mapping}
\end{figure}

\noindent\textbf{MatMul Implementation \& Mapping.}
Fig. \ref{mapping}(b) illustrates the execution of a 2×2 MatMul within an optical core. The column elements of matrix W are programmed into the MRs of each arm, with every arm representing one column of W. The input matrix X is applied row by row to a VCSEL driver, which converts each row vector into light intensities generated by a VCSEL array. These multi-wavelength signals enter the MR bank, where the MRs adjust their intensities at resonance wavelengths to perform optical dot-product computations. The weighted outputs are summed by balanced photodetectors (BPDs), each producing one element of the resulting matrix per computation cycle. Subsequent rows of X are processed in the same manner.
A major challenge in implementing MatMul in ViTs is the large size of weight matrices, often reaching hundreds in dimension, which makes mapping to a single optical core impractical. To address this, the matrices are divided into smaller sub-blocks, and multiplication is performed over multiple cycles. As shown in Fig.~\ref{mapping}(a), input vectors are segmented and sequentially applied to the corresponding weight sub-blocks. In the test design, the VCSEL array generates 32 wavelengths per cycle, enabling 32 parallel optical signals distributed across 64 waveguide arms for simultaneous multiplication with stored weights. Partial results are accumulated each cycle, and the final output is obtained by summing these results. This mapping strategy maximizes wavelength and spatial parallelism, fully exploiting the optical core’s computational capacity.

\vspace{-2mm}
\subsection{Noise Sources \& Deployment Scenario}
\textbf{Noise Modeling.}
We consider three dominant noise sources in MR–based photonic accelerators. Although originating from different stages, these impairments all appear as multiplicative perturbations to the ideal computation, degrading MatVec/MatMul accuracy.
\textit{Fabrication variability} arises from geometric deviations in the MRs, producing resonance wavelength shifts. This static mismatch is modeled as $w_i^{\text{actual}} = w_i(1+\epsilon_i)$ with $\epsilon_i \sim \mathcal{N}(0,\sigma_{\text{fab}}^2)$, where $\sigma_{\text{fab}} = \sigma_\lambda / \Delta\lambda_{\text{FWHM}}$ captures the normalized resonance spread.
\textit{Thermal crosstalk} stems from heater-induced lateral diffusion that perturbs neighboring rings, modeled as $w_i^{\text{actual}} = w_i(1+\epsilon_i+\eta_i)$ with $\eta_i \sim \mathcal{N}(0,\sigma_{\text{thermal}}^2)$.
\textit{Laser fluctuation} affects input amplitude stability. Intensity variations follow $x_i^{\text{actual}} = x_i(1+\zeta)$, where $\zeta \sim \mathcal{N}(0,\sigma_{\text{laser}}^2)$ and may represent global or per-channel fluctuations.
Together, these three noise sources define the effective operating envelope of MR-based accelerators and motivate the need for noise-aware training and compensation strategies.

 \begin{figure}[t] \vspace{-1em}
  \centering
\includegraphics[width=0.99\linewidth]{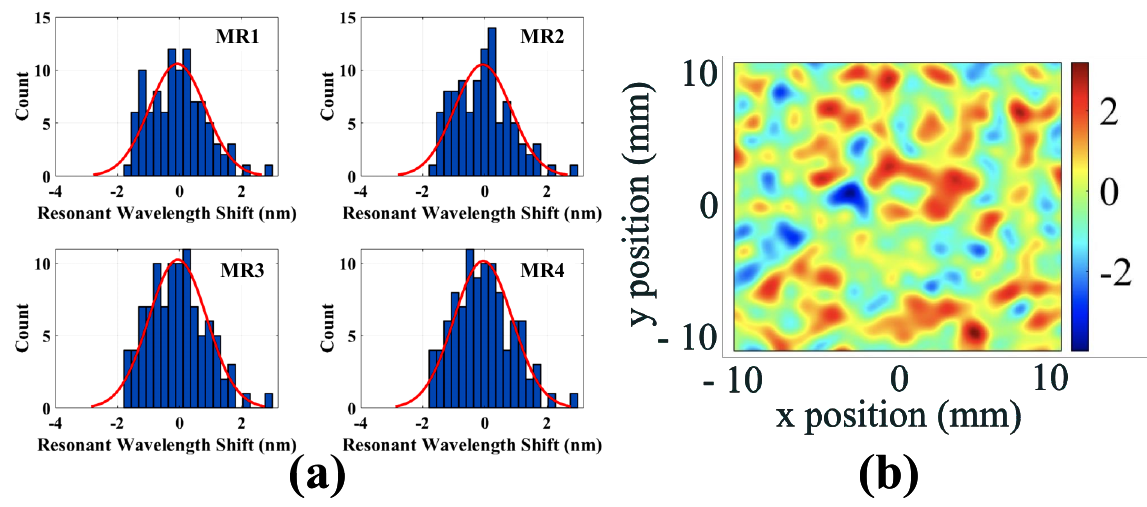} \vspace{-1.0em}
\caption{\small (a) Resonance wavelength shift distribution of a 4 randomly selected MR bank placed at 100 random locations on the variation map, where each data point represents the shift observed in an individual MR at a specific location. (b) Variation heat map showing spatial process-induced deviations across the chip used for MR placement analysis.}
  \label{variation} \vspace{-2.1em}
 \end{figure}

\noindent
\textbf{Fabricated Device Trimming.}
In \textit{pre-trimming} noise modeling, fabrication variability is treated as the dominant stochastic factor. MRs are highly sensitive to nanometer-scale geometric deviations, producing resonance shifts of $\sim$1~nm. Normalized by the linewidth ($\Gamma \approx 1.2$~nm), this yields large multiplicative noise ($\sigma \approx 0.8$), representing worst-case behavior before calibration. To evaluate realistic variation, we generated a virtual variation map (correlation length $l_w=1$~mm). The MR bank with 15 MRs was placed at 100 random locations of this map, and the resonance shift distribution was recorded (Figs.~\ref{variation}(b)). SOI thickness variation is omitted. While our fabricated MRs support 4-bit resolution, we also include 8- and 32-bit projections for potential future capability.
In the \textit{post-trimming} regime, process-induced offsets are treated as deterministic DC errors corrected via wafer-scale trimming \cite{hagan2019post}. Recent studies show resonance alignment within $\pm 32$~pm \cite{jayatilleka2021post}, reducing variability by over an order of magnitude. Post-trim modeling therefore replaces large zero-mean errors with small residual jitter and a systematic MAC-level bias. Remaining noise sources include: (1) residual jitter ($\sigma_\lambda \sim 50{-}100$~pm), giving $\sigma_{ring} \approx \sigma_\lambda/\Gamma \approx 0.05{-}0.1$; (2) slow thermal/operational drift; and (3) fast stochastic noise (laser RIN, readout noise).
To mitigate these imperfections, \textit{two-stage tuning} is used. Thermo-optic (TO) trimming provides coarse, permanent alignment, while electro-optic (EO) tuning, applied periodically, offers fine real-time correction for drift and short-term fluctuations. This complementary TO–EO strategy ensures manufacturability and stability across large MR arrays \cite{sunny2021crosslight}.

\noindent
$\textbf{Signed Weight Mapping.}$ The \textit{balanced differential encoding} is adopted to improve system robustness and enable signed weight representation. Each signed weight \( w_s \in [-1,1] \) is encoded using a pair of unipolar MRs as \( w^{(+)} = (w_s + 1)/2 \) and \( w^{(-)} = (1 - w_s)/2 \), ensuring \( w^{(+)} + w^{(-)} = 1 \) and \( w_s = w^{(+)} - w^{(-)} \). This constant-sum encoding maintains nearly uniform optical power across channels, thereby reducing common-mode fluctuations induced by laser or thermal variations. The resulting quantized levels \( w^{(\pm)} \in [0,1] \) are mapped to detuning values via the inverse Lorentzian relation \( L(\Delta) = 1 / [1 + (2\Delta / \Gamma)^2] \) and \( \Delta = (\Gamma / 2)\sqrt{1/L - 1} \), with \( \Gamma = 1.2~\text{nm} \) denoting the FWHM linewidth. Detuning values are clamped at \( \Delta_{\max} = 2.2~\text{nm} \) to ensure linearity and avoid operating in the Lorentzian tails.
In our modeling framework, the \textit{post-trimming regime} is explicitly incorporated. Wafer-scale trimming aligns each MR’s resonant wavelength \( \lambda_0 \) with residual spread within \( \pm 32~\text{pm} \). The LUT entries for detunings \( (\Delta^{(+)}, \Delta^{(-)}) \) are therefore defined relative to the trimmed \( \lambda_0 \), ensuring consistent reference for both training and inference. Residual stochastic variations are modeled as \( \Delta^{(\pm)}_{\text{actual}} = \Delta^{(\pm)}_{\text{LUT}} + \delta\lambda \), where \( \delta\lambda \sim \mathcal{N}(\mu_r, \sigma_\lambda^2) \), \( \sigma_\lambda \in [20,40]~\text{pm} \) captures the measured post-trim jitter, and \( \mu_r \) represents small systematic biases.

When aggregated across multiple MRs in a MAC operation, the cumulative impact of independent noise sources can be expressed as \( Y_{\text{noisy}} = \sum_{i=1}^{N} x_i (1 + \zeta_i) w_i (1 + \epsilon_i + \eta_i) \), yielding an expected error \( \delta Y = \sum x_i w_i (\zeta_i + \epsilon_i + \eta_i) \) and total output variance \( \text{Var}(\delta Y) = \sum x_i^2 w_i^2 (\sigma_{\text{laser}}^2 + \sigma_{\text{fab}}^2 + \sigma_{\text{thermal}}^2) \). Accordingly, the relative multiplicative noise on the MAC result follows \( \epsilon_{\text{MAC}} = \delta Y / Y \sim \mathcal{N}\!\left(0, \frac{\sum x_i^2 w_i^2 (\sigma_{\text{laser}}^2 + \sigma_{\text{fab}}^2 + \sigma_{\text{thermal}}^2)}{(\sum x_i w_i)^2}\right) \). This expression quantitatively links device-level imperfections to system-level inference degradation, forming the basis for robust noise-aware training and hardware–software co-optimization.

 \begin{figure}[t] \vspace{-1em}
  \centering
\includegraphics[width=0.99\linewidth]{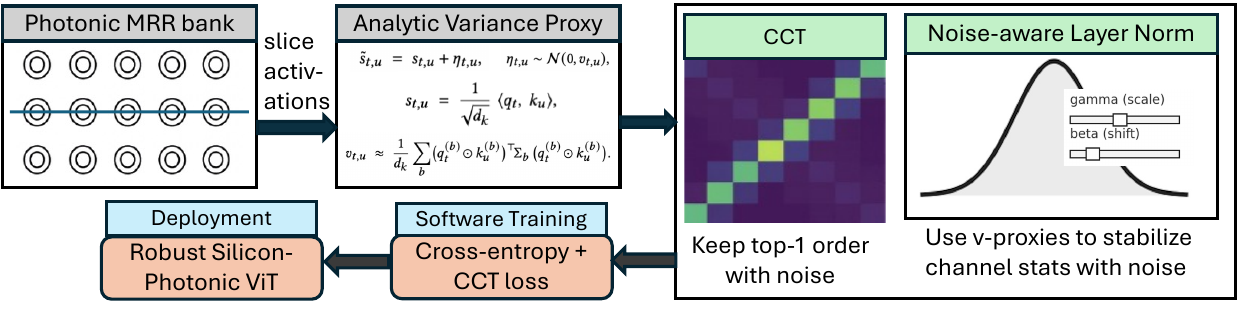} \vspace{-1.0em}
\caption{\small End-to-end noise-aware ViT for photonic hardware. Measured microring-bank statistics inform closed-form variance proxies, driving two algorithmic defenses: chance-constrained attention (CCT) to preserve ranking under noise, and noise-aware LayerNorm for FFN channel stability. Training jointly minimizes task and consistency losses; at deployment, the same weights yield robustness to fabrication, thermal, and laser noise on the photonic core.}
  \label{cct+ln} \vspace{-2em}
 \end{figure}

\vspace{-3mm}
\section{Proposed Training}

\noindent\textbf{Chance-Constrained Training}: Modern ViTs route context by comparing attention logits within each query row. In photonic execution, these logits are realized by analog matrix–vector products whose outputs are perturbed by bank–level variability and runtime fluctuations. Cross–entropy training maximizes likelihood at the class output but leaves the \emph{pairwise orderings} of attention logits unconstrained. Because the softmax allocation is a monotone function of these orderings, small perturbations that flip the top–1 key for a query token can redirect context and cascade through subsequent layers. We aim to directly control the probability of such flips under a measured noise model, converting “robust in expectation” into a \emph{probabilistic margin} requirement at the locus where noise matters most. Consider one head and one query position \(t\). The clean attention logits are $s_{t,u} \;=\; \frac{1}{\sqrt{d_k}}\;\langle q_t,\;k_u\rangle$, with \(q_t,k_u\in\mathbb{R}^{d_k}\) produced by the learned projections. On our hardware, each vector is routed through an array of microring banks; we denote by \(q_t^{(b)},k_u^{(b)}\in\mathbb{R}^{15}\) the slices traversing bank \(b\). Measured per–bank statistics provide either a variance \(\sigma_b^2\) or a covariance \(\Sigma_b\in\mathbb{R}^{15\times 15}\) that captures fabrication spread and thermal fluctuations at that bank. Under the standard small–noise regime for analog MACs, the perturbed logit admits a zero–mean Gaussian approximation
\[
\tilde s_{t,u} \;=\; s_{t,u} + \eta_{t,u}, 
\qquad 
\eta_{t,u}\sim\mathcal{N}(0,v_{t,u}),
\]
where the variance proxy \(v_{t,u}\) is computed analytically from activations and bank statistics without Monte–Carlo sampling. When only per–bank variances are available we use
\[
v_{t,u}\;\approx\; \frac{1}{d_k}\sum_{b}\sigma_b^2\,\big\|q_t^{(b)}\big\|_2^2\,\big\|k_u^{(b)}\big\|_2^2,
\]
and when per–bank covariances are available we tighten this to $v_{t,u}\;\approx\; \frac{1}{d_k}\sum_{b}\!\big(q_t^{(b)}\!\odot k_u^{(b)}\big)^\top\!\Sigma_b\,\big(q_t^{(b)}\!\odot k_u^{(b)}\big)$. Both forms are differentiable in \(q_t^{(b)}\) and \(k_u^{(b)}\) and therefore in the learnable projections.

Let \(i^*(t)=\arg\max_{u} s_{t,u}\) denote the clean top–1 key in row \(t\), and let \(m_{t,ij}=s_{t,i^*}-s_{t,j}\) be the clean margin against a competitor \(j\neq i^*\). Assuming independent per–logit perturbations (a conservative modeling choice that upper–bounds the flip probability in the presence of mild positive correlations), the noisy margin is \(\tilde m_{t,ij}=m_{t,ij}+\delta_{t,ij}\) with \(\delta_{t,ij}\sim\mathcal N(0,\sigma_{t,ij}^2)\) and $\sigma_{t,ij}^2 \;=\; v_{t,i^*}+v_{t,j}$. The probability that noise reverses the ordering is then
\[
\Pr\!\big[\tilde m_{t,ij}\le 0\big] \;=\; \Phi\!\left(-\frac{m_{t,ij}}{\sigma_{t,ij}}\right),
\]
with \(\Phi\) the standard normal CDF. This expression highlights the relevant quantity: the \emph{variance–normalized} margin \(m/\sigma\). Rather than indirectly influencing this ratio through generic augmentation, we enforce a target confidence \(\tau\in(0,1)\) by requiring
\begin{equation}
\frac{m_{t,ij}}{\sigma_{t,ij}} \;\ge\; z_\tau \quad\text{for important competitors } j,\ \ z_\tau=\Phi^{-1}(\tau)
\end{equation}
Because hard constraints would stall optimization, we employ a convex hinge surrogate aggregated over a small adversarial set \(\mathcal N_K(t)\) of the top competitors in row \(t\) (e.g., the next largest logits and those with largest \(v_{t,u}\)), where $ [x]_+=\max(x,0)$:
\[
\mathcal L_{\mathrm{CCT}}
\;=\;
\frac{1}{B}\sum_{t}\sum_{j\in\mathcal N_K(t)}
\left[z_\tau - \frac{s_{t,i^*}-s_{t,j}}{\sqrt{v_{t,i^*}+v_{t,j}}}\right]_+\]

The complete training objective augments task cross–entropy with this chance–constraint penalty (see Fig. \ref{cct+ln}),
\[
\min_{\theta}\ \ \mathbb{E}_{(x,y)}\!\left[
\mathcal L_{\mathrm{CE}}\big(f_\theta(x),y\big)\;+\;\lambda_{\mathrm{CCT}}\ \mathcal L_{\mathrm{CCT}}
\right],
\]
where \(\theta\) denotes all network parameters and \(\lambda_{\mathrm{CCT}}>0\) balances accuracy and robustness. Gradients back–propagate through the variance proxies into \(q_t^{(b)}\) and \(k_u^{(b)}\), encouraging the model to reshape its internal representations so that attention mass does not concentrate on bank–slices that inflate \(v_{t,u}\). Because \(v_{t,u}\) is computed once per forward pass from slice norms (or Hadamard products under \(\Sigma_b\)), the added overhead is linear in the number of tokens and heads and does not introduce sampling variance.

We observe that two implementation details improve stability without altering the formulation. First, we use a curriculum on \(\tau\), beginning with a moderate confidence (e.g., \(\tau=0.80\), \(z_\tau\approx 0.84\)) and annealing to a stringent target (e.g., \(\tau=0.95\), \(z_\tau\approx 1.645\)) as the classifier saturates; this mirrors margin–based curricula in robust optimization and avoids over–regularizing early epochs. Second, we restrict \(\mathcal L_{\mathrm{CCT}}\) to a subset of layers and heads where attention is known to be most semantically critical (early global and late refinement blocks), which reduces compute and focuses the constraint where flips are most harmful. When per–bank covariances are available, replacing the diagonal proxy with the covariance form tightens \(\sigma_{t,ij}\) and further reduces conservatism, although the training code path is identical. The chance–constrained loss provides an interpretable robustness guarantee: for every query row \(t\) and every competitor \(j\in\mathcal N_K(t)\), the trained model maintains the intended ordering with probability at least \(\tau\) under the measured bank–level Gaussian envelope. Unlike undifferentiated noise injection, which equalizes perturbations across irrelevant and decisive pairs, the proposed loss concentrates capacity on preserving the few pairwise relations that govern the softmax allocation. Because the construction is analytic and differentiable, it integrates seamlessly with standard supervised fine–tuning; robustness is encoded in the weights through the normalized attention margins that directly determine photonic behavior.

\noindent\textbf{Noise-Aware Layer Normalization}: The perturbations induced by our SiPh-based system alter the empirical statistics of hidden activations, breaking the implicit assumption in standard Layer Normalization (LN) that observed feature variance faithfully reflects the underlying signal distribution. As a result, the inflated variance caused by additive noise leads to over-normalization, where meaningful feature contrast is suppressed and inter-layer variance amplification is exacerbated. Formally, for an input feature vector $\mathbf{x} \in \mathbb{R}^d$, conventional LN computes:
\vspace{-2mm}
\[
\hat{\mathbf{x}} = \frac{\mathbf{x} - \mu}{\sqrt{\sigma^2 + \epsilon}},
\quad
\mu = \frac{1}{d} \sum_{i=1}^{d} x_i,
\quad
\sigma^2 = \frac{1}{d} \sum_{i=1}^{d} (x_i - \mu)^2.
\]
Under noise $\mathbf{n} \sim \mathcal{N}(0, \sigma_n^2)$, the observed activation becomes
$\tilde{\mathbf{x}} = \mathbf{x} + \mathbf{n}$,
yielding the empirical variance $
\tilde{\sigma}^2 = \frac{1}{d} \sum_{i=1}^{d} (\tilde{x}_i - \tilde{\mu})^2 = \sigma^2 + \sigma_n^2$, where $\sigma_n^2$ represents the expected variance introduced by device fluctuations.
When $\tilde{\sigma}^2$ is used for normalization, the scaling denominator $\sqrt{\tilde{\sigma}^2 + \epsilon}$ becomes larger than necessary, effectively compressing the true signal amplitude and allowing noise to dominate the normalized representation. To miitgate this distortion, we introduce a \textit{Noise-Aware LayerNorm (NALN)} (see Fig. \ref{cct+ln}) that corrects the normalization scale using a noise-aware variance estimator: $
\hat{\mathbf{x}}_{\text{NALN}} =
\frac{\tilde{\mathbf{x}} - \tilde{\mu}}
{\sqrt{\max(\tilde{\sigma}^2 - \sigma_n^2, 0) + \epsilon}}$, where $\sigma_n^2$ is a noise variance proxy obtained from noise model.
By subtracting the expected noise variance, NALN disentangles structural feature variability from stochastic fluctuations.
This operation can be viewed as an \textit{unbiased variance correction} under the assumption that
$\mathbf{n}$ and $\mathbf{x}$ are independent:
\[
\mathrm{Var}[\tilde{\mathbf{x}}] = \mathrm{Var}[\mathbf{x}] + \mathrm{Var}[\mathbf{n}],
\quad
\Rightarrow
\mathrm{Var}[\mathbf{x}] \approx \mathrm{Var}[\tilde{\mathbf{x}}] - \sigma_n^2.
\]
From an optimization standpoint, NALN mitigates the layer-to-layer amplification of normalization noise. Because standard LN couples variance estimates across layers, even mild perturbations can propagate as multiplicative scaling errors, leading to gradient instability and biased updates.
By stabilizing the normalization scale, NALN reduces stochastic variance in both forward and backward passes, yielding smoother convergence and improved generalization under noisy or quantized conditions.




\section{Experimental Results}
\noindent\textbf{Noise injection and setup.}
We inject hardware-realistic noise into all ViT linear layers (Q/K/V and output projections, FFN layers) and the MHSA attention-score computation to emulate photonic MAC imperfections. Q and V are perturbed by weight noise from fabrication and thermal variations ($\sigma_{\mathrm{fab}}$, $\sigma_{\mathrm{thermal}}$), while K and attention logits—corresponding to input-driven optical signals—are perturbed by input noise ($\sigma_{\mathrm{laser}}$). The same noise model is applied to all linear and convolutional operators so every layer contains both weight and input noise. We fix $\sigma_{\mathrm{thermal}}{=}0.1$ and $\sigma_{\mathrm{laser}}{=}0.05$, and vary $\sigma_{\mathrm{fab}}\in[0.05,0.4]$ (measurements indicate $[0.05,0.1]$, but we extend to 0.4 for all tasks and up to 0.7 on CIFAR-10 as a stress test). We evaluate four configurations: (i) clean baseline, (ii) noisy model with/without standard fine-tuning, (iii) CCT-only fine-tuning to isolate NALN, and (iv) joint CCT+NALN. We use ViT-Tiny/Small on CIFAR-10 (224×224), ViT-Base on TinyImageNet (224×224), and ViT-Base within Mask R-CNN for dense prediction tasks, including COCO object detection and segmentation. Results are reported as mean and best top-1 accuracy for classification, and average precision (AP) metrics for detection/segmentation, averaged over ten noisy inference runs.

\begin{figure}[t]
    \centering
    \includegraphics[width=1\linewidth]{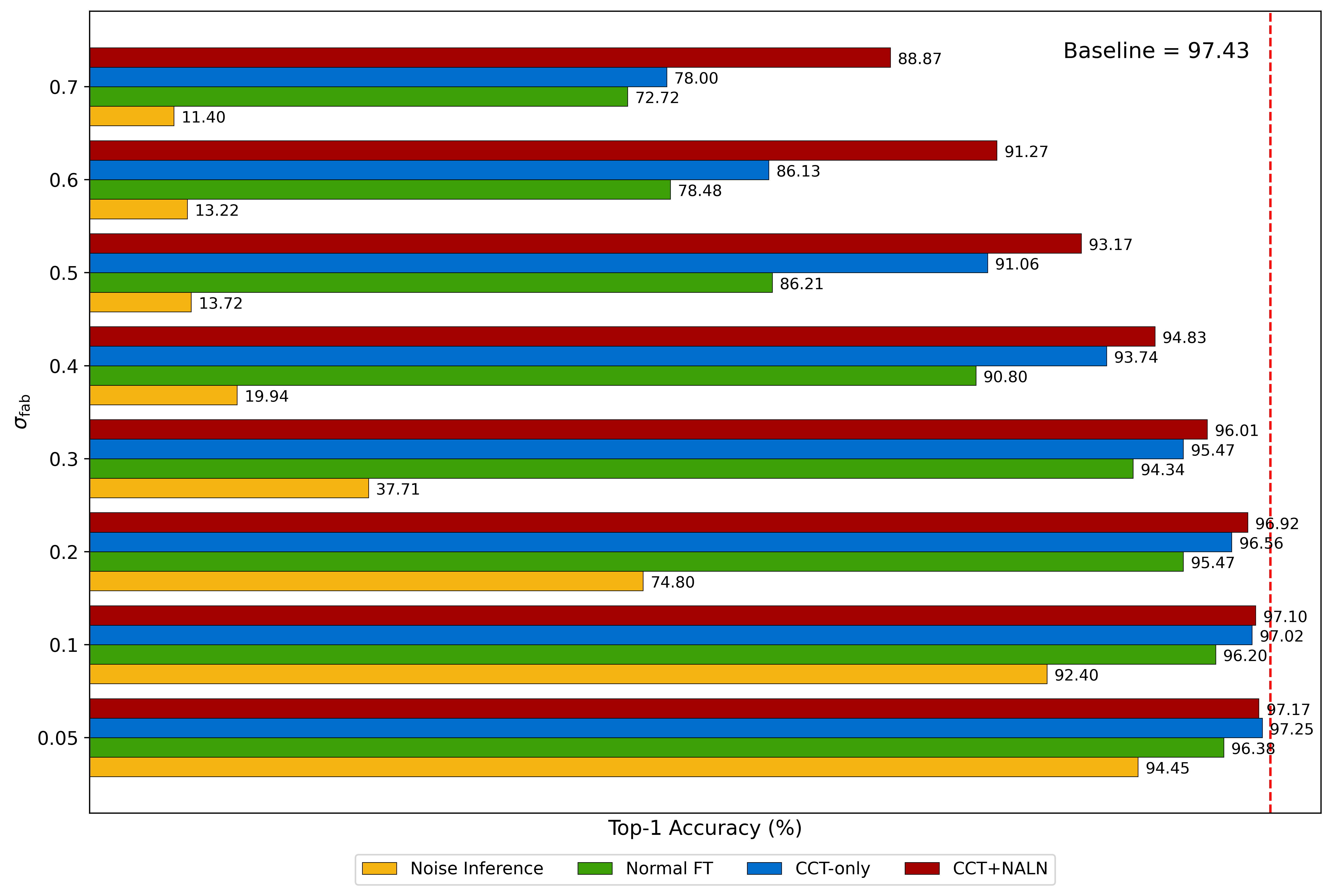}
    \vspace{-1em}
    \caption{Top-1 mean accuracy (\%) of \textbf{ViT-Tiny on CIFAR-10} under different fabrication noise levels $\sigma_{\mathrm{fab}}$. Results are shown for direct noisy inference, normal fine-tuning, CCT fine-tuning, and CCT+NALN.}
    \vspace{-1em}
    \label{fig:vit-tiny-bar}
\end{figure}

\noindent\textbf{Classification Tasks:}
Across fabrication-noise levels in Fig.~\ref{fig:vit-tiny-bar}, we observe a consistent improvement in robustness moving from normal fine-tuning to CCT and finally to CCT+NALN. At moderate noise ($\sigma_{\mathrm{fab}} = 0.20$), direct noisy inference exhibits a severe accuracy drop, while normal fine-tuning recovers much of the loss. CCT provides an additional boost, and CCT+NALN achieves the higher 96.92\% accuracy, nearly matching clean-model performance. As noise increases, these differences become more pronounced. For example, at $\sigma_{\mathrm{fab}} = 0.50$, direct inference remains heavily degraded, normal fine-tuning yields only partial recovery, whereas CCT and CCT+NALN deliver substantial gains, with the latter consistently outperforming all variants. Even at the extreme $\sigma_{\mathrm{fab}} = 0.70$ level—where direct inference collapses—CCT+NALN still restores accuracy to roughly 89\%. Overall, while standard fine-tuning provides some resilience, CCT and especially CCT+NALN offer much stronger robustness under photonic-hardware noise. Besides ViT-Tiny, the improvement are also clearly reflected at lower noise levels in ViT-Small, as shown in Table~\ref{tab:vit-small-results}. For instance, at $\sigma_{\mathrm{fab}} = 0.20$, direct noisy inference drops to 93.75\%, while normal fine-tuning improves mean accuracy to 96.32\%. In contrast, CCT+NALN further elevates performance to 96.51\% (mean) and 96.81\% (best). This improvement demonstrates that CCT+NALN not only compensates for the degradation caused by fabrication noise but also delivers higher robustness compared to standard fine-tuning. 

\begin{table}[!t]
\caption{\small Top-1 accuracy (\%) of ViT-Small on CIFAR-10 under different fabrication noise levels 
$\sigma_{\mathrm{fab}}$.}
\vspace{-1em}
\centering
\scalebox{0.9}
{\begin{tabularx}{0.5\textwidth}{>{\centering\arraybackslash}p{1.2cm} >{\centering\arraybackslash}p{1cm} >{\centering\arraybackslash}p{1.6cm} >{\centering\arraybackslash}p{1.8cm} >{\centering\arraybackslash}p{1.8cm}}
\toprule
\textbf{Model (Dataset)} & \textbf{$\boldsymbol{\sigma_{\mathrm{fab}}}$} & \textbf{Noise Inference} & \textbf{Normal FT (Mean / Best)} & \textbf{CCT+NALN (Mean / Best)} \\
\midrule
\multirow{5}{*}{\parbox{1.5cm}{\centering ViT-Small (CIFAR-10)}}
 & 0    & 97.91 & -             & -            \\
 & 0.05 & 97.25 & 97.31 / 97.41 & 97.55 / 97.83 \\
 & 0.10 & 96.78 & 96.87 / 97.07 & 97.48 / 97.62 \\
 & 0.20 & 93.75 & 96.32 / 96.55 & 96.51 / 96.81 \\
 & 0.40 & 50.70 & 90.78 / 91.25 & 91.40 / 91.80 \\
\bottomrule
\end{tabularx}}
\vspace{-2em}
\label{tab:vit-small-results}
\end{table}

For Tiny-ImageNet classification (Table~\ref{tab:tinyimagenet_results}), we fine-tune ViT-Base with CCT+NALN for 100 epochs using AdamW, adjusting the learning rate according to noise severity. The model remains highly stable under mild fabrication noise: at $\sigma_{\mathrm{fab}} = 0.05$, accuracy drops by only about one percentage point from the clean baseline, and CCT+NALN yields consistent improvements. With stronger noise, accuracy degradation becomes more visible, yet the fine-tuning procedure recovers a substantial portion of the loss—for instance, pushing accuracy at $\sigma_{\mathrm{fab}} = 0.20$ back into 84.03\%. These results indicate that noise-aware adaptation remains effective even for more complex, higher-resolution tasks.

\noindent\textbf{Object Detection and Segmentation}: Consistent with the classification setup, noise is injected only into the optical-domain backbone, while the feature pyramid and detection heads remain noise-free to isolate backbone perturbations. Models are trained for 20 epochs on $224\times224$ inputs using AdamW with a three-stage learning-rate decay from $10^{-4}$ to $10^{-6}$. As shown in Table~\ref{table:coco_rotated}, increasing fabrication noise leads to clear degradation in COCO detection and segmentation accuracy: detection AP drops from 42.18 to 39.70, 38.42, and 32.26. With CCT+NALN fine-tuning, AP recovers to 40.30, 39.57, and 37.01. The same trend holds for AP$^{50}$, AP$^{75}$, and AP$^{s/m/l}$. Segmentation exhibits a similar pattern, with AP falling from 37.88 to 35.62, 34.50, and 28.92, and recovering to 36.23, 35.66, and 33.35 after fine-tuning. These results show that CCT+NALN consistently mitigates performance loss across all evaluated noise levels.

\begin{table}[!t]
\caption{\small Top-1 accuracy (\%) of ViT-Base on Tiny-ImageNet on different fabrication noise levels 
$\sigma_{\mathrm{fab}}$.}
\vspace{-1em}
\centering
\scalebox{0.8}{
\begin{tabularx}{0.5\textwidth}{>{\centering\arraybackslash}p{1cm} >{\centering\arraybackslash}p{2cm} >{\centering\arraybackslash}p{2.5cm} >{\centering\arraybackslash}p{2.5cm}}
\toprule
\textbf{Model (Dataset)} & \textbf{$\boldsymbol{\sigma_{\mathrm{fab}}}$} & \textbf{Noise Inference} & \textbf{CCT+NALN (Mean / Best)} \\
\midrule
\multirow{4}{*}{\parbox{1.5cm}{\centering ViT-Base (Tiny-ImageNet)}}
 & 0 & 86.14 & - \\
 & 0.05 & 85.16 &  85.58/85.71   \\
 & 0.10 & 84.84 &  85.24/85.42   \\
 & 0.20 & 82.49 &  83.72/84.03   \\
 & 0.40 & 65.89 &  79.50/79.84   \\
\bottomrule
\end{tabularx}}
\vspace{-5.5mm}
\label{tab:tinyimagenet_results}
\end{table}

\begin{table}[t]
\caption{\small COCO detection and segmentation AP on different $\sigma_{\mathrm{fab}}$. 
Values are Without/With CCT+NALN fine-tuning.}
\vspace{-1em}
\centering
\setlength{\tabcolsep}{4pt}
\scalebox{0.9}{
\begin{tabular}{lcccc}
\toprule
\textbf{Metric} & 0 & 0.05  & 0.10 & 0.20  \\
\midrule
\multicolumn{5}{l}{\textit{Object detection}} \\
AP      & 42.18 & 39.70 / 40.30 & 38.42 / 39.57 & 32.26 / 37.01 \\
AP$^{50}$ & 62.69 & 59.65 / 60.60 & 58.07 / 59.83 & 50.17 / 56.74 \\
AP$^{75}$ & 46.07 & 43.26 / 43.94 & 41.72 / 42.98 & 34.73 / 40.10 \\
AP$^s$  & 21.89 & 19.71 / 20.58 & 18.85 / 20.03 & 15.00 / 18.33 \\
AP$^m$  & 45.90 & 43.03 / 43.45 & 41.54 / 42.72 & 34.75 / 39.60 \\
AP$^l$  & 56.05 & 53.37 / 53.92 & 51.83 / 52.89 & 43.52 / 50.00 \\
\midrule
\multicolumn{5}{l}{\textit{Instance segmentation}} \\
AP      & 37.88 & 35.62 / 36.23 & 34.50 / 35.66 & 28.92 / 33.35 \\
AP$^{50}$ & 59.66 & 56.60 / 57.50 & 55.08 / 56.79 & 47.26 / 53.72 \\
AP$^{75}$ & 40.53 & 37.99 / 38.63 & 36.63 / 38.02 & 30.35 / 35.33 \\
AP$^s$  & 15.51 & 13.77 / 14.37 & 13.07 / 14.06 & 10.05 / 12.55 \\
AP$^m$  & 40.29 & 37.53 / 38.16 & 36.25 / 37.59 & 29.93 / 34.75 \\
AP$^l$  & 56.92 & 54.14 / 54.91 & 52.59 / 54.23 & 44.60 / 51.42 \\
\bottomrule
\end{tabular}}
\vspace{-5mm}
\label{table:coco_rotated}
\end{table}

\begin{figure}[b] \vspace{-1em}
\centering
\includegraphics [height=3.5cm,width=0.99\linewidth]{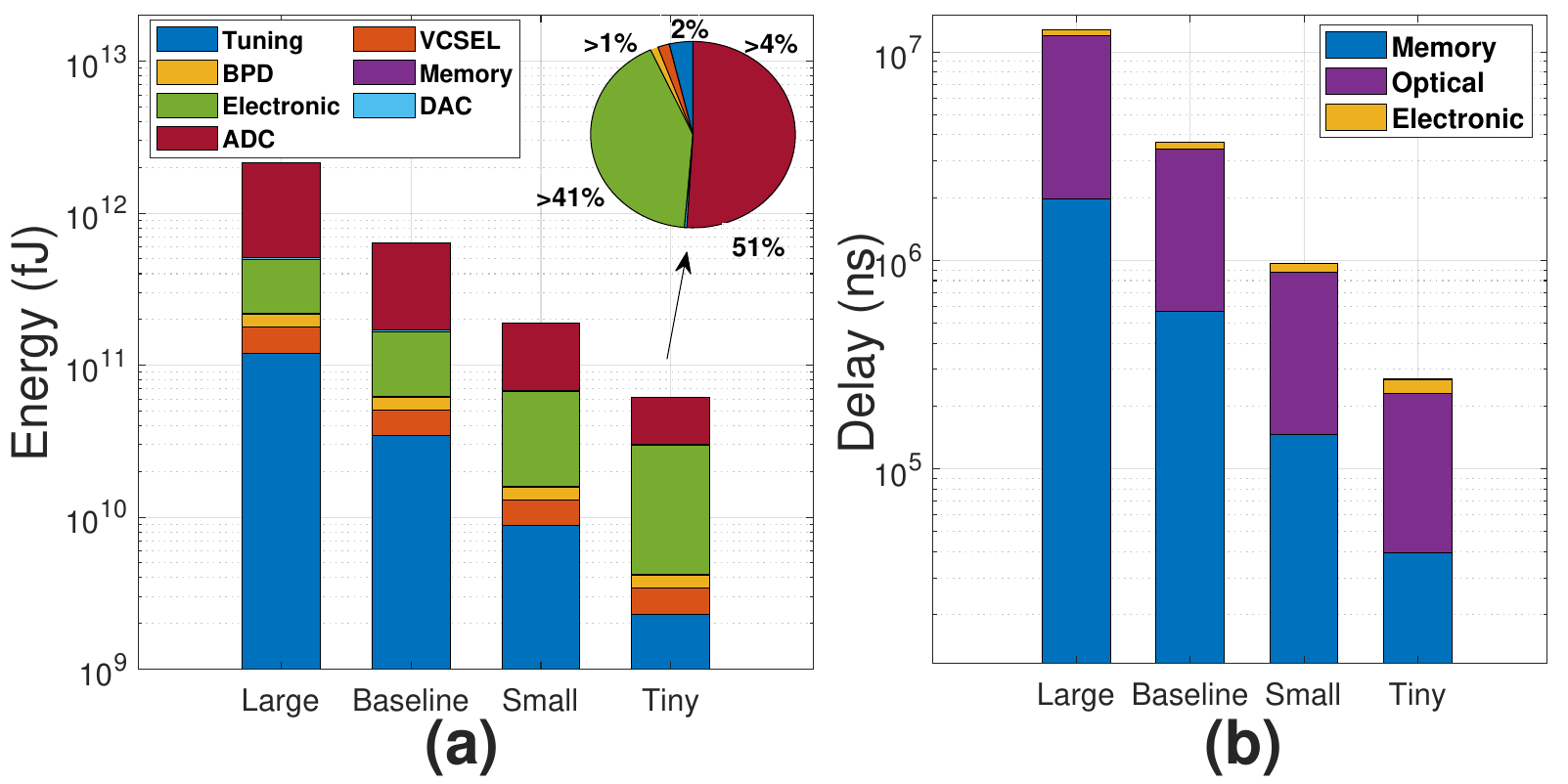}
\vspace{-1.5em}
\caption{(a) Processing energy and (b) delay breakdown for various ViT models processed with input image size 224$\times$224.} 
\vspace{-1.2em}
\label{edp}
\end{figure}
\noindent\textbf{Performance Breakdown.} To assess the performance of the under-test architecture, both energy consumption and processing latency were analyzed across four transformer models—Large, Baseline, Small, and Tiny—using 224×224 input images. As illustrated in Fig. \ref{edp} (a), the total energy is distributed among Tuning, VCSEL, BPD, ADC, DAC, memory, and electronic processing units, showing a clear reduction trend for smaller networks. Despite the primary computation being executed in the optical analog domain, the pie chart for the Tiny-224×224 case reveals that ADCs dominate overall energy consumption, emphasizing the need to further shift processing toward the analog domain to minimize data-conversion overhead.
The corresponding latency analysis in Fig. \ref{edp} (b) demonstrates that optical processing, including ADC and DAC operations, accounts for the majority of the total delay, as it handles most of the transformer computation. As mentioned, the main contributor to the overall energy consumption is the ADC. For runtime trimming to compensate for noise, only the tuning block (including the DAC and tuning circuits) needs to be utilized. This portion accounts for less than 5\% of the total energy and delay in the system.
As explained in Section 3.2, the post-fabrication EO compensation is applied only after several tuning iterations, introducing approximately a 20\% overhead in energy and delay when performed once every five iterations. Overall, the total overhead for noise compensation remains below 1\% of the total energy and delay budget. Considering this negligible overhead, we still achieve approximately 9\% higher accuracy for the case of $\sigma_{fab}$ = 0.7 according to Fig.~\ref{fig:vit-tiny-bar}.

\begin{figure} \vspace{-1em}
\centering
\includegraphics [width=1.01\linewidth]{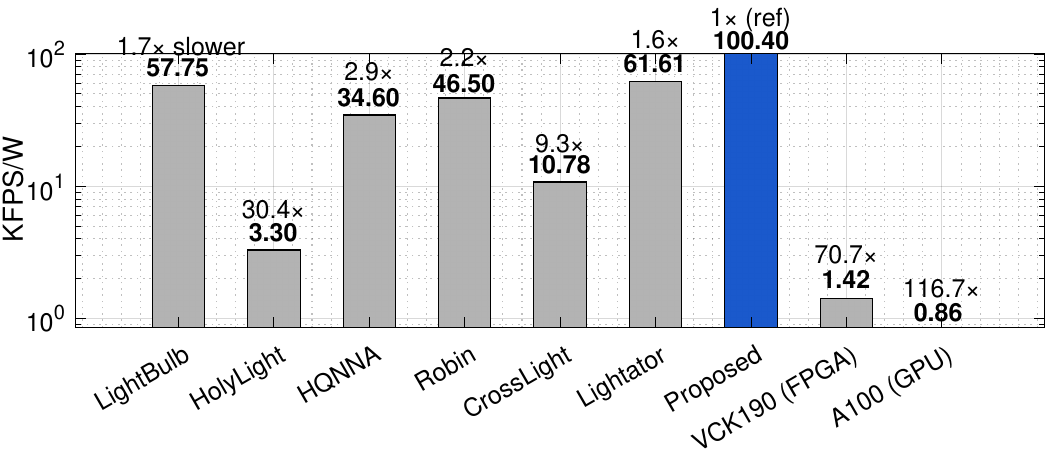}
\vspace{-2.6em}
\caption{\small Performance (KFPS/W) comparison with baseline SiPh designs and VCK190 FPGA and NVIDIA A100 GPU.} 
\vspace{-1.5em}
\label{cmp}
\end{figure}

\noindent\textbf{KFPS/W Comparison.}
To quantify the benefits of the proposed design, we evaluate its energy efficiency against two state-of-the-art electronic inference platforms—the Xilinx VCK190 FPGA and the NVIDIA A100 GPU with TensorRT—following the protocol in \cite{dong2024eq}. All systems process the same ViT model in INT8 format, ensuring fair comparison across low-precision hardware backends. As shown in Fig.~\ref{cmp}, the optical accelerators exhibit a striking advantage, outperforming electronic baselines by two to three orders of magnitude. The proposed design achieves a peak efficiency of 100.4~KFPS/W, whereas the VCK190 reaches only 1.42~KFPS/W ({$70.6\times$ slower}) and the A100 delivers 0.86~KFPS/W ({$116.7\times$ slower}). Fig.~\ref{cmp} also compares several MR-based optical accelerators, including LightBulb~\cite{zokaee2020lightbulb}, HolyLight~\cite{liu2019holylight}, HQNNA~\cite{sunny2022silicon}, Robin~\cite{sunny2021robin}, CrossLight~\cite{sunny2021crosslight}, Lightator~\cite{morsali2024lightator}, and the proposed design. Because most architectures were not originally developed for ViT workloads, each was reconstructed using our simulator under a consistent area budget (20–60$,\text{mm}^2$). The proposed design delivers $100.4$~KFPS/W and outperforms LightBulb ({$1.7\times$ slower}), HolyLight ({$30.4\times$ slower}), HQNNA ({$2.9\times$ slower}), Robin ({$2.2\times$ slower}), and CrossLight ({$9.3\times$ slower}); only Lightator trails modestly by {$1.6\times$}.

\noindent\textbf{Side-by-Side Comparison with GPU \& FPGA.}
The comparison in Table~\ref{vit_latency_energy} shows the substantial latency advantages of SiPh accelerators over electronic platforms for ViT inference. Even after scaling FPGA and GPU latency values by an additional order of magnitude to account for pipeline, buffering, and scheduling overheads, SiPh still achieves more than an order-of-magnitude improvement across all ViT model sizes. The benefit is most notable for the Small and Base variants, where SiPh reduces inference time by up to $116\times$ and $45\times$ relative to FPGA and GPU implementations. Owing to the optical domain’s parallelism and low propagation delay, SiPh maintains low latency as model complexity grows, whereas electronic architectures suffer from memory access bottlenecks and interconnect congestion. Energy results show similar trends: although FPGA is competitive for the Tiny model, deeper pipelines and off-chip memory traffic lead to up to $6.7\times$ higher energy for Large ViT. SiPh remains efficient because optical MACs require negligible incremental energy. \vspace{-1em}

\begin{table}[h]
\centering
\caption{Inference Time and Energy for ViT Variants.} \vspace{-1.2em}
\scalebox{0.6}{
\begin{tabular}{l|ccc|ccc}
\hline
& \multicolumn{3}{c|}{\textbf{Latency ($\mu$s)}} 
& \multicolumn{3}{c}{\textbf{Energy ($\mu$J)}} \\
\textbf{Model} 
& \textbf{FPGA (VCK190)} & \textbf{GPU (A100)} & \textbf{SiPh} 
& \textbf{FPGA (VCK190)} & \textbf{GPU (A100)} & \textbf{SiPh} \\
\hline
Tiny  &
27429 ($\times$102$\uparrow$) &
10528 ($\times$39.1$\uparrow$) &
269 &
54.86 ($\times$0.89$\downarrow$) &
263.19 ($\times$4.29$\uparrow$) &
61.4 \\
Small &
108050 ($\times$112.2$\uparrow$) &
41472 ($\times$43.1$\uparrow$) &
963 &
216.1 ($\times$1.15$\uparrow$) &
1036.8 ($\times$5.51$\uparrow$) &
188 \\
Base  &
428880 ($\times$116.9$\uparrow$) &
164610 ($\times$44.8$\uparrow$) &
3670 &
857.8 ($\times$1.35$\uparrow$) &
4115.3 ($\times$6.46$\uparrow$) &
637 \\
Large &
151010 ($\times$11.8$\uparrow$) &
579620 ($\times$45.3$\uparrow$) &
12800 &
3020.3 ($\times$1.40$\uparrow$) &
14490 ($\times$6.74$\uparrow$) &
2150 \\
\hline
\end{tabular}}
\label{vit_latency_energy}
\end{table}

\vspace{-1.4em}
\section{Conclusion} \vspace{-0.2em}
In this work, we introduce Light-Bound Transformers, a unified framework for deploying ViTs on silicon-photonic accelerators by incorporating device constraints and hardware noise into training and inference. Using noise-aware attention and normalization, we recover near clean-model accuracy across vision tasks under strong analog noise and tight energy budgets. Experiments on simulation and hardware-in-the-loop setups show up to two orders of magnitude energy gains over digital accelerators with minimal accuracy loss, without requiring in-situ learning or hardware modifications.

\bibliographystyle{unsrt}
\bibliography{Reference}

\end{document}